\documentclass[11pt, letterpaper]{article}

\usepackage{ifthen}
\newcommand{\Bset}[1]{\setboolean{#1}{true}}
\newcommand{\Bunset}[1]{\setboolean{#1}{false}}

 \usepackage{amssymb}
 \usepackage{amsmath}  
 \usepackage{amsthm}  
 \usepackage{latexsym}
 \usepackage{amsfonts}
 \usepackage{psfrag}   
 \usepackage{url}   

  \usepackage[dvips]{graphicx}
  \usepackage[usenames,dvipsnames]{color}

\newcommand{\Bif}[3]{\ifthenelse{\boolean{#1}}{#2}{#3}}

\newtheoremstyle{remark}{}{}{}{}{\itshape}{:}{ }{\thmname{#1}\thmnumber{ #2}}
\newtheoremstyle{remslide}{}{}{}{}{\itshape}{:}{ }{\thmname{#1}}

   {\theoremstyle{remark}}

\newcounter{tmp_id_cnt}
\newcommand{\nospell}[1]{#1}  

\newcommand{\mydef}[2]{\def#1{#2}}

\newcommand{\newident}[3][*]{\ifthenelse{\equal{*}{#1}}
 {\newcommand{#2}[1][*]
  {\ifthenelse{\equal{*}{##1}}
   {\nospell{\mbox{\Ensuremath{{\mathit{#3}}}}}}
   {\ifthenelse{\equal{b}{##1}}
     {\nospell{\mbox{\Ensuremath{{\mathbf{#3}}}}}}
     {#3}}}}
 {\mydef{#2}{#3}}}

\newcommand{\newmat}[3][*]{\ifthenelse{\equal{*}{#1}}
 {\newcommand{#2}[1][*]
  {\ifthenelse{\equal{b}{##1}}
   {\nospell{\mbox{\Ensuremath{\mathbf{#3}}}}}
   {\ifthenelse{    \( \equal{*}{##1} \and \not \boolean{in_math_mode} \)
    \or \( \not \equal{*}{##1} \and \boolean{in_math_mode} \)}
    {\nospell{\mbox{\Ensuremath{#3}}}}
    {#3}}}}
 {\mydef{#2}{#3}}}

\newcommand{\MyMakeTheoMacros}[3]{ \newcommand{#2}[2][]{\ifthenelse{\equal{}{##1}}
  {\begin{#1} ##2 \end{#1}}
  {\begin{#1}\label{##1} ##2\end{#1}}}
 \newcommand{#3}[3][]{\ifthenelse{\equal{}{##1}}
  {\begin{#1}{\bf\e{##2}} ##3 \end{#1}}
  {\begin{#1}{\bf\e{##2}}\label{##1} ##3\end{#1}}}}

\newcommand{\MyMakeDupTheoMacros}[6]{ \MyMakeTheoMacros{#1}{#2}{#3}
 \newcommand{#4}[3]{  \begin{#1}\label{##1} ##2\end{#1}}
 \newcommand{#6}[2]{\def\my_tmp_id{my_tmp_id_\arabic{tmp_id_cnt}}
  \newtheorem*{\my_tmp_id}{#5~\ref{##1}}
  \begin{\my_tmp_id} ##2 \end{\my_tmp_id}\stepcounter{tmp_id_cnt}}}

\newcommand{\MyMakeRefMacros}[3]{\newcommand{#1}[2][]
 {\ifthenelse{\equal{}{##1}}{#2~\ref{##2}}{#3~\ref{##1} and~\ref{##2}}}}

\newcommand{\MyMakeEqRefMacros}[3]{\newcommand{#1}[2][]
 {\ifthenelse{\equal{}{##1}}{#2~\eqref{##2}}{#3~\eqref{##1} and~\eqref{##2}}}}

\newcommand{\abstr}[1]{	\begin{abstract}
		#1
	\end{abstract}}

  \newcommand{\bibentry}[8]{\bibitem[\nospell{#8}]{#1} {\textup #3}. 
  \ifthenelse{\equal{}{#6}}
   {\newblock \textrm{#4.} \newblock {\em #5}, #7.}
   {\newblock \textrm{#4.} \newblock {\em #5, #6}, #7.}}

  \newcommand{\inputbib}{\bibentry{A04_Lim}{Aaronson}{S. Aaronson}{Limitations of Quantum Advice and One-Way Communication}{Proceedings of the 19th IEEE Conference on Computational Complexity}{pp. 320-332}{2004}{A04}

\bibentryPerCom{H. Buhrman}{B}

\bibentry{BFL90}{Babai, Fortnow and Lund}{L. Babai, L. Fortnow and C. Lund}{Non-Deterministic Exponential Time Has Two-Prover Interactive Protocols}{Proceedings of the 31st Annual Symposium on Foundations of Computer Science}{pp.  16-25}{1990}{BFL90}

\bibentry{BJK04_Exp}{Bar-Yossef, Jayram and Kerenidis}{Z. Bar-Yossef, T. S. Jayram and I. Kerenidis}{Exponential separation of quantum and classical one-way communication complexity}{Proceedings of 36th Symposium on Theory of Computing}{pp. 128-137}{2004}{BJK04}

\bibentry{CHTW04_Cons}{Cleve, H{\o}yer, Toner and Watrous}{R. Cleve, P. H{\o}yer, B. Toner and J. Watrous}{Consequences and limits of nonlocal strategies}{Proceedings of the 19th IEEE Conference on Computational Complexity}{pp. 236-249}{2004}{CHTW04}

\bibentry{GKRW06_Boun}{Gavinsky, Kempe, Regev and de Wolf}{D. Gavinsky, J. Kempe, O. Regev and R. de Wolf}{Bounded-error Quantum State Identification and Exponential Separations in Communication Complexity}{Proceedings of the 38th Symposium on Theory of Computing}{}{2006}{GKRW06}

\bibentry{GS86}{Goldwasser and Sipser}{S. Goldwasser and M. Sipser}{Private Coins versus Public Coins in Interactive Proof Systems}{Proceedings of the 18th Symposium on Theory of Computing}{pp. 59-86}{1986}{GS86}

\bibentry{JRS05_Pri}{Jain, Radhakrishnan and Sen}{R. Jain, J. Radhakrishnan and P. Sen}{Prior Entanglement, Message Compression and Privacy in Quantum Communication}{Proceedings of the 20th IEEE Conference on Computational Complexity}{pp. 285-296}{2005}{JRS05}

\bibentry{KM03_Qua}{Kobayashi and Matsumoto}{H. Kobayashi and K. Matsumoto}{Quantum Multi-prover Interactive Proof Systems with Limited Prior Entanglement}{Journal of Computer and System Sciences 66(3)}{pp. 429-450}{2003}{KM03}

\bibentry{KSW04_Qua}{Klauck, Spalek and de Wolf}{H. Klauck, R. Spalek and R. de Wolf}{Quantum and Classical Strong Direct Product Theorems and Optimal Time-Space Tradeoffs}{Proceedings of the 45th Annual Symposium on Foundations of Computer Science}{pp. 12-21}{2004}{KSW04}

\bibentry{N91_Pri}{Newman}{I. Newman}{Private vs. Common Random Bits in Communication Complexity}{Information Processing Letters 39(2)}{pp. 67-71}{1991}{N91}

\bibentry{NS96_Pub}{Newman and Szegedy}{I. Newman and M. Szegedy}{Public vs. Private Coin Flips in One Round Communication Games}{Proceedings of the 28th Symposium on Theory of Computing}{pp. 561-570}{1996}{NS96}

\bibentry{R95_A_par}{Raz}{R. Raz}{A parallel repetition theorem}{Proceedings of the 27th Symposium on Theory of Computing}{pp. 447-456}{1995}{R95}

\bibentry{S05_Ten}{Shi}{Y. Shi}{Tensor norms and the classical communication complexity of bipartite quantum measurements}{Proceedings of the 37th Symposium on Theory of Computing}{}{2005}{S05}}

  \newcommand{\bib}[1][]{   \begin{thebibliography}{ABCDE00}
     \frenchspacing
     \inputbib
     #1
   \end{thebibliography}}

  \newcommand{\citePerCom}{\cite}
  \newcommand{\bibentryPerCom}[2]{   \bibitem[\nospell{#2}]{#1} {\textup #1} - \textit{Personal communication}.}

\MyMakeTheoMacros{fact}{\fct}{\nfct}

\MyMakeRefMacros{\fctref}{Fact}{Facts}

 \MyMakeDupTheoMacros{lemma}{\lem}{\nlem}{\lemdup}{Lemma}{\lemrep}

\MyMakeRefMacros{\lemref}{Lemma}{Lemmas}

\MyMakeTheoMacros{corollary}{\crl}{\ncrl}

\MyMakeRefMacros{\crlref}{Corollary}{Corollaries}

\MyMakeTheoMacros{proposition}{\prp}{\nprp}

\MyMakeRefMacros{\prpref}{Proposition}{Propositions}

 \MyMakeDupTheoMacros{claim}{\clm}{\nclm}{\clmdup}{Claim}{\clmrep}

\MyMakeRefMacros{\clmref}{Claim}{Claims}

\MyMakeDupTheoMacros{theorem}{\theo}{\ntheo}{\theodup}{Theorem}{\theorep}

\MyMakeRefMacros{\theoref}{Theorem}{Theorems}

\MyMakeTheoMacros{definition}{\defi}{\ndefi}

\MyMakeRefMacros{\defiref}{Definition}{Definitions}

\MyMakeTheoMacros{problem}{\prob}{\nprob}

\MyMakeRefMacros{\probref}{Problem}{Problems}

\MyMakeTheoMacros{protocol}{\prot}{\nprot}

\MyMakeRefMacros{\protref}{Protocol}{Protocols}

\newcommand{\prf}[2][]{\ifthenelse{\equal{}{#1}}
 {\begin{proof}\renewcommand{\qedsymbol}{$\blacksquare$} #2 \end{proof}}
 {\begin{proof}[Proof of #1]
  \renewcommand{\qedsymbol}{$\blacksquare_{\mbox{\it{\scriptsize{#1}}}}$}
  #2 \end{proof}}}

\newcommand{\sect}[2][]{\ifthenelse{\equal{}{#1}}
 {\section{#2}}
 {\section{#2}\label{#1}}}

\newcommand{\ssect}[2][]{\ifthenelse{\equal{}{#1}}
 {\subsection{#2}}
 {\subsection{#2}\label{#1}}}

\newcommand{\sssect}[2][]{\ifthenelse{\equal{}{#1}}
 {\subsubsection{#2}}
 {\subsubsection{#2}\label{#1}}}

\MyMakeRefMacros{\chref}{Chapter}{Chapters}

\MyMakeRefMacros{\sref}{Section}{Sections}

\MyMakeRefMacros{\ssref}{Subsection}{Subsections}

\MyMakeRefMacros{\sssref}{Subsection}{Subsections}

\newboolean{in_math_mode}
\setboolean{in_math_mode}{false}

\newcommand{\IfMathMode}{\Bif{in_math_mode}}

\newcommand{\MathModeOn}{\Bset{in_math_mode}}

\newcommand{\MathModeOff}{\Bunset{in_math_mode}}

\newcommand{\Ensuremath}[1]{\IfMathMode  
 {\ensuremath{#1}}
 {\MathModeOn\ensuremath{#1}\MathModeOff}}

\newcommand{\fbr}[1]{\IfMathMode  
 {$#1$}                             
 {\MathModeOn$#1$\MathModeOff}}

\newcommand{\fnbr}[1]{\mbox{\fbr{#1}}}  

\newcommand{\fla}[2][*]{\ifthenelse{\equal{}{#1}} {\fbr{#2}} {\fnbr{#2}} }

\newcommand{\mat}[2][]{\ifthenelse{\equal{}{#1}}  
 {\begin{displaymath} \MathModeOn
  #2
  \MathModeOff \end{displaymath}    }{  \begin{equation} \MathModeOn \label{#1}
  #2
  \MathModeOff \end{equation}}}

\newcommand{\matal}[2][]{\mat[#1]{\begin{split} #2 \end{split}}}

\newcommand{\f}{\fla}

\newcommand{\m}{\mat}

\newcommand{\mal}{\matal}

\newcommand{\twocase}[4]
 {\begin{cases}#1 &\txt{#2}\\#3 &\txt{#4}\end{cases}}

\MyMakeEqRefMacros{\equref}{Equation}{Equations}

\MyMakeEqRefMacros{\expref}{Expression}{Expressions}

\MyMakeEqRefMacros{\inequref}{Inequality}{Inequalities}

\newcommand{\bracref}[1]{(\ref{#1})}

\newcommand{\bref}{\bracref}

\MyMakeRefMacros{\figref}{Figure}{Figures}

\newcommand{\poly}{\mathop{\mbox{poly}}}
\newcommand{\tr}{\mathop{\mbox{tr}}}

\newcommand{\h}[2][]{\mathop{\mathbf{H}}_{#1}\left[{#2}\right]}

\newcommand{\E}[2][]{\mathop{\mathbf{E}}_{#1}\left[{#2}\right]}

\newcommand{\PR}[2][]{\mathop{\mathbf{Pr}}_{#1}\left[{#2}\right]}

\newcommand{\U}[1][]{\ifthenelse{\equal{}{#1}}
 {{\cal U}}
 {{\cal U}_{#1}}}

\newcommand{\pl}[1][]{\nospell{\ifthenelse{\equal{}{#1}}
 {\mbox{-s}}
 {\fla{#1}\mbox{-s}}}}

\newcommand{\ord}[1][]{\nospell{\ifthenelse{\equal{}{#1}}
 {\mbox{'th}}
 {\ifthenelse{\equal{1}{#1}}{$1$\mbox{'st}}{\ifthenelse{\equal{2}{#1}}{$2$\mbox{'nd}}{\ifthenelse{\equal{3}{#1}}{$3$\mbox{'rd}}{\fla{#1}\mbox{'th}}}}}}}

\newmat[]{\llp}{\left(}
\newmat[]{\rrp}{\right)}

\newmat[]{\lla}{\left\langle}
\newmat[]{\rra}{\right\rangle}

\newmat[]{\dt}{\cdot}
\newmat[]{\tm}{\cdot}
\newmat[]{\xor}{\oplus}
\newmat[]{\sbseq}{\subseteq}
\newmat[]{\eps}{\varepsilon}
\newmat[]{\deq}{\stackrel{\textrm{def}}{=}}

\newmat{\NN}{\mathbb{N}}
\newmat{\RR}{\mathbb{R}}

\newcommand{\fr}[3][*]{\ifthenelse{\equal{}{#1}}
 {#2/#3}
 {\frac{#2}{#3}}}

\newcommand{\set}[2][]{\ifthenelse{\equal{}{#1}}
 {\Ensuremath{\left\{#2\right\}}}
 {\Ensuremath{\left\{#2\left|~#1\right.\right\}}}}

\newcommand{\Max}[2][]{\ifthenelse{\equal{}{#1}}
 {\Ensuremath{\max\left\{#2\right\}}}
 {\Ensuremath{\max_{#1}\left\{#2\right\}}}}

\newcommand{\newfunction}[2]{ \newcommand{#1}[2][*]{\ifthenelse{\equal{*}{##1}}
  {\Ensuremath{#2\llp ##2 \rrp}}
  {#2(##2)}}}

\newfunction{\asO}{O}
\newfunction{\astO}{\tilde O}
\newfunction{\aso}{o}
\newfunction{\asOm}{\Omega}
\newfunction{\astOm}{\tilde \Omega}
\newfunction{\asom}{\omega}
\newfunction{\asT}{\Theta}

\newcommand{\ket}[1]{\Ensuremath{\left|#1\rra}}
\newcommand{\kbra}[1]{\Ensuremath{\left|#1\rra\hspace{-3.5pt}\lla #1\right|}}

\newmat[]{\01}{\set{0,1}}

\newcommand{\sz}[2][]{\ifthenelse{\equal{}{#1}}
 {\Ensuremath{\left|#2\right|}}
 {\Ensuremath{\left|#2\right|_{#1}}}}

\newcommand{\fn}{\footnote}

\newcommand{\nin}{\not\in}  

 \newcommand{\e}{\emph}

\newcommand{\txt}[1]{\textrm{#1}}  

\newcommand{\tb}{\quad}

 \usepackage{hyphenat}  

\date{}
\newident{\NEXP}{NEXP}
\newident{\EXP}{EXP}
\newident{\MIPe}{MIP^*_{e(n)}}

\newident{\HMP}{HMP}
\newident{\Hm}{\HMP_m}
\newident{\Hmk}{\HMP^{(k)}_m}

\title{On the Role of Shared Entanglement}

\author{  {\bf Dmitry Gavinsky} \\
  {\small Department of Computer Science}\\
  {\small University of Calgary}\\
  {\small Calgary, Alberta, Canada, T2N 1N4}\\}

\begin{document}

\maketitle

\thispagestyle{empty}

\abstr{Despite the apparent similarity between shared randomness and shared
entanglement in the context of Communication Complexity, our understanding of
the latter is not as good as of the former.
In particular, there is no known ``entanglement analogue'' for the famous
theorem by Newman, saying that the number of shared random bits required
for solving any communication problem can be at 
most logarithmic in the input length (i.e., using more than $\asO[]{\log n}$
shared random bits would not reduce the complexity of an optimal solution).

In this paper we prove that the same is not true for entanglement.
We establish a wide range of tight (up to a polylogarithmic factor)
entanglement vs.\ communication tradeoffs for relational problems. 
The low end is:\ for any $t>2$, reducing shared entanglement
from $log^tn$ to $\aso[]{log^{t-2}n}$ qubits can increase the communication
required for solving a problem almost exponentially, from $\asO[]{log^tn}$
to $\asOm[]{\sqrt n}$. 
The high end is:\ for any $\eps>0$, reducing shared entanglement from $n^{1-\eps}\log n$ to $\aso[]{n^{1-\eps}/\log n}$ can increase the required communication from $\asO[]{n^{1-\eps}\log n}$ to $\asOm[]{n^{1-\eps/2}/\log n}$.
The upper bounds are demonstrated via protocols which are \e{exact} and work
in the \e{simultaneous message passing model}, while
the lower bounds hold for \e{bounded-error protocols}, even
in the more powerful \e{model of 1-way communication}.
Our protocols use shared EPR pairs while the lower bounds apply
to any sort of prior entanglement.

We base the lower bounds on a strong direct product theorem for communication complexity of a certain class of relational problems.
We believe that the theorem might have applications outside the scope of this work.}

\newpage
\setcounter{page}{1}

\sect{Introduction}
Suppose that Alice, Bob and Charlie play the following game: Alice receives
an $n$-bit binary string $x$, Bob receives a string $y$ of the same length,
they both send some information to Charlie, who then tries to guess (based on
the received messages) whether $x=y$ or not.
The goal is for Alice and Bob 
to send as short messages as possible, such that Charlie
would still be able to answer correctly with probability at least $3/4$.
Assume that before the game starts Alice and Bob choose two random binary
strings $r_1$ and $r_2$, of length $2^n$ each.
Then they treat their $n$-bit inputs as indices in the range $[1..2^n]$ and
send to Charlie
the bits which are on the positions $x$ and $y$ of $r_1$ and $r_2$ (i.e.,
Alice and Bob send 2 bits each).
Eventually, Charlie decides that $x=y$ if the pairs of bits received from
Alice and Bob are the same, otherwise he declares that $x\ne y$.
It is clear that if Charlie guesses that $x=y$ then he is correct with
probability $3/4$ and if he says that $x\ne y$ then he is certainly right.
So, we see that the problem can be solved by communicating only $4$ bits
(for any input length $n$).
On the other hand, Newman and Szegedy \cite{NS96_Pub} have shown that if Alice
and Bob do not share random bits then they must communicate at least
\asOm{\sqrt n} bits in order to win the game with any constant probability
greater than $1/2$.
 
We can let our players use the laws of quantum mechanics in order to further
increase their strength.
Specifically, they can share \e{entanglement}.\fn
{Note that even if Alice and Bob share a quantum state the communication
channels are still assumed to be classical.
For simplicity in this paper we do not deal with quantum communication, even
though it seems that some of our results generalize to that case.}
In this case they are allowed to apply any quantum-mechanical
operation to their subspaces of the common Hilbert space.
In particular, they can perform measurements and their
behavior may depend on the outcomes of the measurements.

If the players share a sequence of random bits (chosen uniformly and
independently) we say that they are using \e{shared} or \e{public
randomness} (also called a \e{public coin}).
If the players share a quantum state we say that they are using
\e{shared entanglement} (note that it would be useless to share a state which
is not entangled w.r.t.\ the players' local subspaces).
It is easy to see that in the model of shared entanglement the players are at
least as strong as they are in the model of shared randomness
($k$ independent shared EPR pairs can be measured locally in
order to get $k$ perfect random bits).

In this paper we will deal with a longstanding open question regarding the
power of quantum entanglement in communication.

\ssect{Shared randomness and shared entanglement}
Let us generalize our framework, suppose that Alice and
Bob have to fulfill some computation-flavored distributed task.
As before, the players are located far from one another, so that communication between them is expensive or even impossible.
The players are all powerful from the computational aspect.

Two well know instances of this framework are 2-prover proof systems
and various models of 2-party communication complexity.

In the first case Alice and Bob are provers, they can communicate with a
\e{verifier} but not with one another.
The verifier is computationally limited.
The goal of the provers is to convince the verifier that some string $x$
belongs to a language $L$, when checking validity of that statement is beyond
the verifier's computationally ability.
If the verifier believes, based on its communication with the provers, that
$x\in L$ then we say that $x$ is \e{accepted}, otherwise it is \e{rejected}.
A language $L$ \e{has a valid 2-prover proof system} if Alice and Bob can make
the verifier accept (with high probability) any $x\in L$, but making it to
accept some $y\nin L$ would be (almost) impossible.

In the models of 2-party communication complexity Alice and Bob receive one
piece of input each, respectively denoted by $x$ and $y$.
In the strongest considered model communication between Alice and Bob is
possible but expensive, it goes in many rounds (first Alice sends a
message to Bob, then Bob replies, then Alice sends another message and so on).
Their goal is to compute (with high probability) some function $f(x,y)$ using
the smallest possible amount of communication.

One possible restriction of the model is \e{1-way communication}: Alice is 
permitted to send a message to Bob, after that he has to produce an output
(based on $y$ and the message from Alice).
Note that unlike the unrestricted case, the 1-way model is not symmetric
w.r.t.\ $x$ and $y$.
Sometimes even more restricted (symmetric) case is considered which has been
described in the beginning:
there is another participant called a \e{referee}, Alice and Bob can send one
message each to the referee and it has to produce an output based on those two
messages.
This model is called \e{simultaneous message passing (SMP)}, it is arguably
the weakest setting of 2-party communication complexity that is still
interesting.

A \e{communication protocol} is a description of the behavior of all
the participants.
The \e{communication cost} of a protocol is the maximum possible total
length of the messages sent according to the protocol till the output is
produced.
The optimization problem is to find a least expensive protocol which enables
the players to solve their task;
the \e{communication cost} of a communication task is the cost of an
optimal protocol.

Sometimes the communication task is defined not as a function but rather as a \e{relational problem}.
In that case for a pair $(x,y)$ in the input there can be defined any number of good answers (no good answer means that the pair can never be given as input).
In this paper we allow this more general form of communication problems.

In all these models (both 2-prover proof systems and
2-party communication complexity settings)
we understand relatively well what the power of shared
randomness is.
In the case of proof systems two classical all-powerful provers
can prove to a polynomially-bounded verifier\fn
{In particular, that means that the communication cost of a proof can be at
most polynomial in the input length.}
membership in $L$ if and only if $L\in\NEXP$ (\cite{BFL90}, \cite{R95_A_par}).
It is also known that shared randomness
does not affect the power of a system (\cite{GS86}).

In the case of 2-party communication complexity the situation is slightly more
complicated.
It has been demonstrated by Newman \cite{N91_Pri} that we can assume
without loss of generality that the number of shared random bits used by a
protocol is at most logarithmic in the input length. 
Therefore, availability of shared randomness cannot reduce significantly the
complexity in the models where Alice sends at least one message to Bob 
(she can append the required number of randomly chosen bits to her message,
that would increase the cost of a protocol only by an additive logarithmic
term).
In the case of SMP the presence of shared randomness can make a difference.
For instance, as mentioned in the beginning, the \e{equality problem} can be
solved by a protocol of constant cost when shared randomness is available,
whereas without shared randomness the complexity 
becomes~\asOm{\sqrt n}~(\cite{NS96_Pub}).

We know much less about the role of shared entanglement in the context of these models.
We do not know what the power of 2-prover proof systems is when the provers share entanglement.\fn
{As far as we know today, such systems can be more powerful, less powerful, or even incomparable to the standard 2-prover systems, since adding power to provers can, in general, help them to establish a true argument as well as to cheat.}
Moreover, Cleve, H{\o}yer, Toner and Watrous \cite{CHTW04_Cons} have shown that the known protocol which accepts \NEXP\ in the standard 2-prover system cannot achieve the same goal in the presence of shared entanglement, unless $\EXP=\NEXP$.
For the restricted case when the provers share only polynomial (in $n$) number
of qubits, it has been demonstrated by Kobayashi and Matsumoto
\cite{KM03_Qua} that only languages from \NEXP\ can be accepted (but again,
maybe not all of them).

In the area of communication complexity very recently a communication task has been found (\cite{GKRW06_Boun}) which can be solved exponentially more efficiently in the SMP model with shared entanglement than in the SMP model with shared randomness (in fact, the problem is equally hard even for 1-way communication with shared randomness).
But it is not known whether any upper bound can be put on the number of qubits in a potentially helpful shared quantum state.

There is a result by Shi \cite{S05_Ten} which says (informally) that adding
large amounts of prior entanglement can reduce the communication no more than
exponentially.
However, Jain, Radhakrishnan and Sen \cite{JRS05_Pri} have shown
that Newman's ``blackbox-type'' proof, which keeps the protocol the same and
just reduces the set of random strings to \asO{n} elements (which in turn can
be represented by \asO{\log n} random bits), cannot be used in order to reduce
the amount of entanglement used.  

Besides, it is not clear whether EPR pairs can be considered as a universal
source of entanglement in the contexts of 2-prover proof systems and 2-party
communication complexity.

\ssect{Our results}
As our main result, we claim that no reasonably sublinear upper bound holds
for the number of potentially useful shared entangled qubits.
Put in contrast to the Newman's theorem, this is a new example of qualitative
difference between the two resources (public coin vs.\ shared entanglement).
Note that our conclusion and that of \cite{GKRW06_Boun} are logically related,
our result can be viewed as a generalization of the models separation in
\cite{GKRW06_Boun}.\fn 
{Assume towards contradiction that shared entanglement is no more powerful
than public randomness (this is a contrapositive to \cite{GKRW06_Boun}).
Then any number of shared qubits can be replaced by similar
number of public random bits, which can be reduced to logarithmic number
(due to Newman) and then simulated by the same number of shared EPR pairs.
So, logarithmic number of shared EPR pairs would always be sufficient,
which is a contrapositive to our main statement.}

Formally, our main result is the following.
\theo[theo_main]{For any monotone increasing function $k(\dt):\NN\to\NN$ there exists a family of relational communication problems such that the problem with input length $n=m\tm k(m)$ can be solved exactly in the SMP model with $k(m)\log(m)$ shared EPR pairs by
a protocol of cost \asO{k(m)\log(m)}.
The same problem requires \asOm{\frac{k(m)\sqrt{m}}{\log m}} communication for its solution with constant-bounded error in the model of 1-way communication with any shared entangled state of \aso{\frac{k(m)}{\log m}} qubits.}

In particular, for any $t>2$ by choosing $k(m)=\log^{t-1}m$ we obtain a problem with input length $n=m\tm\log^{t-1}m$ which can be solved using (less than) $log^tn$ EPR pairs and \asO{log^tn} communication but requires \asOm{\sqrt n} communication with \aso{log^{t-2}n} shared entanglement.
Therefore, limiting the amount of shared entanglement even to super-logarithmic values can result in almost exponential increase in communication cost of a problem.

Alternatively, for any $\eps>0$ choosing $k(m)=m^{\fr[]1\eps-1}$ gives a problem with input length $n=m^{\fr[]1\eps}$, solvable with (less than) $n^{1-\eps}\log n$ EPR pairs and  \asO{n^{1-\eps}\log n} communication but demanding \asOm{\frac{n^{1-\eps/2}}{\log n}} communication with \aso{\frac{n^{1-\eps}}{\log n}} entanglement.
Therefore, no reasonably sublinear upper bound on the number of useful shared qubits can be put.

Note that our protocols use shared EPR pairs whereas the lower bounds hold for any sort of entanglement.
Because our analysis is tight up to a polylogarithmic factor, it can be concluded that for the families of relations we consider it is the case that independent EPR pairs are as good as any sort of shared entanglement can be (up to a polylogarithmic factor).
Our protocols are exact and work in the SMP model, while the lower bounds hold for bounded-error protocols, even in the model of 1-way communication.

Our proof consists of two parts.
First, we establish a strong direct product theorem for a class of relational problems in the model of 1-way communication.\fn
{We call a direct product result \e{strong} if the amount of available resources scales up as the number of instances grows.}
In particular, the theorem is applicable to the relation \HMP\ defined by Bar-Yossef, Jayram and Kerenidis \cite{BJK04_Exp}.
We think that the theorem might be of independent interest.
For instance, the fact that it gives a strong direct product result for the one-way complexity of \HMP\ looks very promising, because this relation, and its modifications, is the only known type of communication problem that demonstrates superpolynomial separation between quantum and classical 1-way models.
Problems based on \HMP\ have been used recently to establish a number of exponential separations between various quantum and classical communication models (cf.\ \cite{BJK04_Exp}, \cite{GKRW06_Boun}).
It might be the case that our strong direct product result can be used to obtain more results based on \HMP.

The second part is a construction of an entanglement-expensive communication task.
We first apply our direct product theorem in order to reduce to
exponentially low the maximum success probability of a protocol which is not
using entanglement.
Then we view a hypothetical protocol which is successful when it
uses a shared entangled state $\rho$ as a distinguisher between $\rho$ and
the maximally mixed state of the same dimension.
If the protocol starts with the maximally mixed shared state then the
players are \e{not entangled}, so the upper bound on success probability
without entanglement must hold.
By the laws of quantum mechanics, any distinguisher between $\rho$ and the
maximally mixed state must be wrong with probability not less than
approximately the inverse of the dimension of the state.
So, our assumption that a protocol starting with $\rho$ is successful leads to
a lower bound on the dimension of $\rho$
(in our case the obtained bounds are tight up to a polylogarithmic factor in terms of the \e{number of qubits} in $\rho$).
Note that the resulting entanglement lower bound itself has the form of a
direct product result (i.e., in order to solve more copies of the original
problem one must accordingly increase the number of shared entangled qubits to
start with).

It can be seen that the described technique is quite general; probably it can
be used in other situations where the ``entanglement complexity'' of a problem
is considered (including all the models we have mentioned). 
For instance, given a corresponding direct product result for the complexity of
a solution without entanglement, the technique can be applied to virtually any
communication complexity model.

We note that the technique of replacing a quantum state under consideration by
the maximally mixed state and upper-bounding the damage caused by such
substitution has been used before in several contexts related to communication
complexity (cf.\ \cite{KSW04_Qua}, \cite{A04_Lim}).
It seems that the technique is quite powerful and might be applicable in
various settings involving quantum mechanics and information processing.

We also apply our entanglement bounding idea in the context of 2-prover proof systems.
We give a partial converse to the result of \cite{KM03_Qua}.
We characterize
the power of 2-prover proof systems, where the provers are allowed to share
entanglement but the number of qubits is bounded by a polynomial \e{fixed a
priori} (i.e., the bound should be a global parameter of the model).\fn
{Note that the model considered by \cite{KM03_Qua} gives the provers more
freedom than we do.
Their provers can use the amount of entanglement which is bounded \e{per
protocol}, while ours are bounded \e{per model}.
It is still open whether the proof system of \cite{KM03_Qua} can accept any
language in \NEXP.}
The power of such proof systems equals \NEXP, i.e., in this case the factor
of entanglement does not affect the power of a system.

\sect{Preliminaries}
In this paper we will deal with relational communication problems.
Formally, a problem will be represented as $P\sbseq X\times Y\times Z$, where
$X=Y=\01^*$ are the sets of inputs to Alice and Bob, correspondingly, and $Z$
is the set of possible answers.
An answer $z\in Z$ is good for input $(x,y)\in X\times Y$ if $(x,y,z)\in P$;
if no such $z$ exists then the combination $(x,y)$ is forbidden
(i.e., it is never given as input).
We will write $X_n$ to denote $X\cap\01^n$ as well as $Y_n$ and $Z_n$ to denote \set[\exists x\in X_n, z\in Z: (x,y,z)\in P]{y\in Y} and \set[\exists x\in X_n, y\in Y: (x,y,z)\in P]{z\in Z}, correspondingly.

Let $P\sbseq X\times Y\times Z$ be a relational problem.
When $P$ is clear from the context, for any $A\sbseq X$, $y\in Y$ and $z\in Z$ we will denote by $A_{|y,z}$ the set \set[(x,y,z)\in P]{x\in A}.

We write $P^k\sbseq X^k\times Y^k\times Z^k$ to address the direct product of $k$ instances of $P$, formally:
\m{P^k=\set[\forall i\in\set{1,..,k}:(x_i,y_i,z_i)\in P]
 {\big((x_1,..,x_k),(y_1,..,y_k),(z_1,..,z_k)\big)},}
in that case we will address $P$ as a \e{single instance} of a problem.
For any $A\sbseq X^k$, $i\le k$, $a_1,..,a_i\in Y$ and $b_1,..,b_i\in Z$ we define: 
\m{A_{|a_1,..,a_i,b_1,..,b_i}\deq\set[\forall 1\le j\le i:(x_j,a_j,b_j)\in P]{x\in A},}
and for $a\in Y$ and $b\in Z$: 
\m{A_{|y_i=a,z_i=b}\deq\set[(x_i,a,b)\in P]{x\in A}.}
Note that in our definitions of $k$-ary direct products we have changed the
natural ordering and grouping of elements, making them more suitable for our
context of communication tasks.

For convenience we assume that 1-way protocols do not use shared randomness (private random bits will be allowed, of course).
As explained earlier, this does not cause any loss of generality because that assumption can, in the worst case, result in adding a logarithmic factor to the communication cost.

For any discrete set $A$ we denote by $\U[A]$ the uniform distribution over $A$.
For a discrete random variable $x$ we denote by $\h{x}$ its Shannon entropy.
Sometimes we write $\h[D]{x}$ for a distribution $D$ to emphasize that $x\sim D$, the same value will be denoted by $\h{D}$ when $x$ is insignificant for the context.

We write $\log$ to denote the logarithmic function with base $2$.

\sect{A strong direct product theorem for relations}
We establish a strong direct product theorem for a class of relational problems in the model of 1-way communication without shared entanglement.

\lem[lem_dir]{Let $P\sbseq X\times Y\times Z$ be a relation.
Let $\sigma(m):\NN\to\RR$ and $\delta(m):\NN\to[0,1]$ be two functions, such that $\log m\le\sigma(m)\le m,$ $\log\llp\frac1{\delta(m)}\rrp\ge4+6\frac{\log(\sz{Z_m})}{\log m}$
and for any distribution $D$ over $X_m$ with $\h[D]{x}\ge m-\sigma(m)$ it holds that
\m{\PR[{(y,z)\sim\U[Y\times Z]}]
 {\PR[{x\sim D}]{(x,y,z)\in P}\ge\frac23}\le\frac{\delta(m)}{\sz{Z_m}}.}
Then for $m\ge64$ and $k\ge\log m$, for any set $B\sbseq (X_m)^k$ of size at least $2^{km-\frac{k\sigma(m)}{\log m}}$ the following holds:\
\m{\PR[{y\sim\U[Y^k]}]
 {\exists z\in Z^k:\sz{B_{|y,z}}\ge(2/3)^{\frac{k}{\log m}}\sz{B}}\le2^{-k}.}}

\prf[\lemref{lem_dir}]{Fix a set $B$ satisfying the lemma condition.
Define:
\m{{\cal E}(y,z)\deq\twocase
 {1}{if $\sz{B_{|y,z}}\ge(2/3)^{\frac{k}{\log m}}\sz{B}$}
 {0}{otherwise},}
and we will use the same notation for the corresponding logical predicate (i.e., ${\cal E}(y,z)$ is satisfied if and only if ${\cal E}(y,z)=1$).
It holds that
\m{\PR[{y\sim\U[Y^k]}]{\exists z\in Z^k:{\cal E}(y,z)}\le
 \E[{y\sim\U[Y^k]}]{\sum_{z\in Z^k}{\cal E}(y,z)}=
 |Z^k|\dt\PR[{(y,z)\sim\U[Y^k\times Z^k]}]{{\cal E}(y,z)}.}

We will upper-bound the expression on the right-hand side.
Before we proceed, let us introduce some notation.
We will address individual coordinates of elements of $X^k$, $Y^k$ and $Z^k$ through $(x_1,..,x_k)$ for an $x\in X^k,$ and similarly for $y\in Y^k$ and $z\in Z^k$.

Let us think of choosing $(y,z)\sim\U[Y^k\times Z^k]$ as a sequential $k$-step process of choosing pairs $(y_i,z_i)\sim\U[Y\times Z],$ not necessarily in the ascending order of \pl[i].
We will specify the order later, so far we denote it by $j_1,..,j_k,$ i.e., at the first step we choose $(y_{j_1},z_{j_1}),$ and so on. 

Let $a_{j_1},..,a_{j_k}$ and $b_{j_1},..,b_{j_k}$ be the choices made for the random variables $y_{j_1},..,y_{j_k}$ and $z_{j_1},..,z_{j_k}$, correspondingly.
Define:\ $B_0\deq B$ and for $1\le i\le k$:\ $B_i\deq {B_{i-1}}_{|y_{j_i}=a_{j_i},z_{j_i}=b_{j_i}}$.

Consider the sequence $\sz{B_0},..,\sz{B_k}$ -- it is monotone non-increasing, and ${\cal E}(y,z)$ exactly means that $\sz{B_k}\ge(2/3)^{k/\log m}\sz{B}$.
Let us say that step $i$ is \e{good} if $\sz{B_i}/\sz{B_{i-1}}\ge2/3$.
Observe that ${\cal E}(y,z)$ occurs only if at least ${k-\frac{k}{\log m}}$ steps were good.
Assuming integer-valued rounding where necessary, we get
\mal{\PR{{\cal E}(y,z)}\le
 &\PR{\txt{at least ${k-\frac{k}{\log m}}$ steps were good}}\le\\
 &{k\choose {k-\frac{k}{\log m}}}\tm
  \Max[i_1,..,i_{k-\frac{k}{\log m}}]
  {\PR{\txt{the steps $i_1,..,i_{k-\frac{k}{\log m}}$ were good}}},}
where the maximum is taken over all $(k-\frac{k}{\log m})$-tuples of pairwise distinct indices from $[k]$.
For the reasons which will become clear later, we do not want to take into consideration the last $\frac{2k}{\log m}$ steps, so we let those steps be good ``for free'' and get
\mal{\PR{{\cal E}(y,z)}\le
 &{k\choose {k-\frac{k}{\log m}}}\tm\Max[i_1,..,i_{k-\frac{3k}{\log m}}]
  {\PR{\txt{the steps $i_1,..,i_{k-\frac{3k}{\log m}}$ were good}}}\le\\
 &2^k\tm\llp\Max[{i\in[k-\frac{2k}{\log m}]}]
  {\PR{\txt{the \ord[i] step was good}}}\rrp^{k-\frac{3k}{\log m}},}
where the first maximum is taken over all $(k-\frac{3k}{\log m})$-tuples of pairwise distinct indices from \f{[k-\frac{2k}{\log m}]}.
We can make our bound tighter using the fact that the event ${\cal E}(y,z)$ implies that all \pl[B_i] are of size at least
$(2/3)^{k/\log m}\sz{B}\ge2^{km-\frac{k\sigma(m)}{\log m}-k}$:\
\mal{&\PR{{\cal E}(y,z)}\le\\
 &\tb2^k\tm\llp\Max[{i\in[k-\frac{2k}{\log m}]}]
  {\PR{\txt{the \ord[i] step was good}\big|\sz{B_{i-1}}\ge
  2^{km-\frac{k\sigma(m)}{\log m}-k}}}\rrp^{k-\frac{3k}{\log m}}.}

Denote:
\m{p_{max}\deq\Max[{i\in[k-\frac{2k}{\log m}]}]
  {\PR{\txt{the \ord[i] step was good}
  \big|\sz{B_{i-1}}\ge2^{km-\frac{k\sigma(m)}{\log m}-k}}}.}
We have seen that
\m[m_adyn]{\PR[{y\sim\U[Y^k]}]{\exists z\in Z^k:{\cal E}(y,z)}\le
 \sz{Z_m}^k\tm2^k\tm p_{max}^{k-\frac{3k}{\log m}}.}

For obtaining the desired bound we will, on each step, try to choose a coordinate which has low chances to give rise to a good step.
Such a coordinate for the \ord[i] step will be chosen adaptively among all those not fixed in the first $i-1$ steps.

Assume that we are at step $i_0$ now, let us see that such a bad coordinate must exists as long as $\sz{B_{i_0-1}}\ge2^{km-\frac{k\sigma(m)}{\log m}-k}$ and $i_0\le k-\frac{2k}{\log m}$.
Let $D_{i_0-1}$ be the uniform distribution over $B_{i_0-1}$, we know that $\h[D_{i_0-1}]{x}\ge km-\frac{k\sigma(m)}{\log m}-k$.
For $j\in [k]$ define $e_j$ to be the entropy of $x_j$ when $x\sim D_{i_0-1}$ and let $J\deq\set[{j\in[k]}]{e_j\ge m-\sigma(m)}.$
Because $\forall j\in [k]: e_j\le m,$ by the pigeonhole principle and entropy subadditivity it must hold that
\mal{ &\sz{J}m+(k-\sz{J})(m-\sigma(m))\ge km-\frac{k\sigma(m)}{\log m}-k\\
 &k-\sz{J}\le\frac{k}{\log m}+\frac{k}{\sigma(m)}\le\frac{2k}{\log m}\\
 &\sz{J}\ge k-\frac{2k}{\log m}.}
Choose arbitrary $j_0\in J,$ such that $y_{j_0}$ and $z_{j_0}$ have not been set yet (it exists because we are only at step $i_0\le k-\frac{2k}{\log m}$).

Observe that
\m{\PR[{(a,b)\sim\U[Y\times Z]}]
 {\sz{{B_{i_0-1}}_{|y_{j_0}=a, z_{j_0}=b}}\ge\frac23\sz{B_{i_0-1}}}
 \le\frac{\delta(m)}{\sz{Z_m}},}
by the theorem assumption about $P$ applied to the \ord[j_0] coordinate of $P^k$.
So we choose $j_0$ as the coordinate to be handled at step $i_0$.
Because $B_i={B_{i-1}}_{|y_{j_0}=a, z_{j_0}=b},$ where $(a,b)\sim\U[Y\times Z],$ we conclude that the probability of \ord[i] step to be good is at most $\delta(m)/\sz{Z_m}.$

Recall that our goal is to upper-bound the value of $p_{max}$.
We have chosen $i_0$ to be any integer not exceeding $k-\frac{2k}{\log m}$, so an upper bound on $\PR{\txt{the \ord[i_0] step was good}}$ under our assumptions is also an upper bound on $p_{max}$.
Therefore, \bref{m_adyn} leads to the required
\m{\PR[{y\sim\U[Y^k]}]{\exists z\in Z^k:{\cal E}(y,z)}\le
 \sz{Z_m}^k\tm2^k\tm
  \llp\frac{\delta(m)}{\sz{Z_m}}\rrp^{k-\frac{3k}{\log m}}\le
 \sz{Z_m}^{\frac{3k}{\log m}}\tm2^k
  \tm(\delta(m))^{k-\frac{3k}{\log m}}\le2^{-k},}
where the last inequality follows from the theorem assumptions regarding $\delta(\dt)$ and $m$.}

The following theorem is straightforward from \lemref{lem_dir}:
\theo[theo_dir]{Let $P\sbseq X\times Y\times Z$ be a relation satisfying the condition of \lemref{lem_dir}.

Then for $m$ large enough and $k\ge\log m$, any 1-way communication protocol of complexity at most $\frac{k\sigma(m)}{\log m}-2$ solves $P^k$ with success probability at most $(2/3)^{\frac{k}{\log m}-2}$.}

\prf[\theoref{theo_dir}]{Assume towards contradiction that there exists a protocol $A$ of complexity at most $\frac{k\sigma(m)}{\log m}-2$ which solves $P^k$ with success probability more than $(2/3)^{\frac{k}{\log m}-2}$.
Then there exists a deterministic protocol $A'$ which does the same when the input distribution is $\U[X^k\times Y^k]$.

Assume that $x\sim\U[X^k]$ and $y\sim\U[Y^k]$.
For any message ever sent by Alice according to $A'$, define its \e{weight} as the probability of the message to be produced and its \e{success} as the probability that $A'$ is successful, conditioned on the message having been produced.
By the pigeonhole principle, with probability at least $3/4$ Alice sends a message of weight at least $1/4$ divided by the number of possible messages.
Similarly, with probability at least $1/3$ Alice sends a message of success at least $2/3$ times the success probability of the protocol.
In other words, there exists some message $\alpha$ such that 
\m{B\deq
 \set[\txt{given $x$, Alice sends $\alpha$ according to $A'$}]{x\in X^k}}
satisfies $\sz{B}\ge2^{\frac{k\sigma(m)}{\log m}},$ and conditioned on Alice sending $\alpha$, Bob is able to produce a correct answer with probability more than $(2/3)^{\frac{k}{\log m}-1}$. 

Let $z(y)$ be defined as the answer produced by Bob according to $A'$, if his own input is $y$ and the message received from Alice is $\alpha$.
Then, according to the previous discussion, it must hold that
\m{\PR[{(x,y)\sim\U[B\times Y^k]}]
 {(x,y,z(y))\in P^k}>(2/3)^{\frac{k}{\log m}-1},}
which leads to
\m{\PR[{y\sim\U[Y^k]}]
 {\PR[{x\sim\U[B]}]{(x,y,z(y))\in P^k}>(2/3)^{\frac{k}{\log m}}}
 >(2/3)^{\frac{k}{\log m}}.}
In other words,
\m{\PR[{y\sim\U[Y^k]}]
 {\sz{B_{|y,z(y)}}\ge(2/3)^{\frac{k}{\log m}}\sz{B}}
 >(2/3)^{\frac{k}{\log m}},}
which contradicts \lemref{lem_dir}.

Our theorem follows.}

\sect{A communication task with an entanglement-expensive solution}
Let $m$ be a power of $2$.
The following relational problem has been first studied by Bar-Yossef, Jayram and Kerenidis \cite{BJK04_Exp}.
\defi{Let $X=\01^m$, and let $M_m$ be the family of all perfect matchings on $m$ nodes, represented as $m/2$-tuples of pairs of vertices connected by an edge.
Then
\m{\Hm=\set[x\in X, y\in M_m, y_a=\set{i,j}]{\big(x,y,(a,x_i\xor x_j)\big)}.}}
In words, Alice receives a binary coloring of $m$ nodes and Bob receives a perfect matching on $m$ nodes; the goal is to say whether a pair of nodes connected by the matchings are colored the same or not.

Let $k$ be an integer greater than $1$.
We define $\Hmk$ as a direct product of $k$ instances of $\Hm$.
\defi{$\Hmk=\set{\big((x_1,\ldots,x_k),(y_1,\ldots,y_k),(z_1,\ldots,z_k)\big)},$ where for
all $i\in\set{1,\ldots,k}$ it holds that \f{(x_i,y_i,z_i)\in\Hm.}}

We will consider the communication complexity of certain sub-families of $\Hmk$ in order to establish our entanglement vs.\ communication tradeoffs.

\ssect[sec_Hm]{Complexity of $\Hm$}
It is known that $\Hm$ can be solved exactly using $\log(m)$ EPR pairs and $\asO{\log(m)}$ bits of communication in the SMP model.
The protocol is a modification of a construction suggested by Buhrman \citePerCom{H. Buhrman} (a similar protocol is used in \cite{GKRW06_Boun}).
For completeness we describe the protocol here.

The starting state of Alice and Bob is
\m{\frac1{\sqrt{m}}\sum_{i\in\01^{\log m}}\ket{i}\ket{i}.}
First, Alice applies phases according to her input $x$:
\m{\frac1{\sqrt{m}}\sum_{i\in\01^{\log m}}(-1)^{x_i}\ket{i}\ket{i}}
and Bob measures with the $m/2$ projectors 
$E_{i,j}=\kbra{i}+\kbra{j}$ induced by the pairs $\set{i,j}\in y$.
After that both players apply a Hadamard transform to each of the $\log n$ qubits of their part of the shared state, which then becomes (ignoring normalization)
\m{\sum_{k,l}\llp(-1)^{x_i+(k\xor l)\cdot i}+(-1)^{x_j+(k\xor l)\cdot j}\rrp
 \ket k\ket l,}
where \set{i,j} is the outcome of the Bob's measurement, $\xor$ denotes the bit-wise xor operation and $\tm$ stands for the inner product $mod~2$ of two vectors.
It follows that $\ket k\ket l$ has non-zero amplitude if and only if 
\m{(k\xor l)\tm(i\xor j)=x_i\xor x_j.}
The players measure the state $\ket k\ket l$ in the computational basis, then Alice sends $k$ and Bob sends $a$, $i$, $j$ and $l$ to the referee, where $y_a=\set{i,j}$.
The referee outputs $\big(a,(k\xor l)\tm(i\xor j)\big),$ and the protocol is always correct.

Concerning the lower bound, it has been demonstrated in \cite{BJK04_Exp} that $\Hm$ is hard for 1-way communication without entanglement.
However, we need a stronger statement, in order to be able to apply \theoref{theo_dir}.
\clm[clm_bjk]{Let $\sigma(m)=\frac{\sqrt{m-1}}{576}$ and $\delta=\frac1{2^{10}}$, then for any distribution $D$ over $X_m$ with $\h[D]{x}\ge m-\sigma(m)$ it holds that 
\m{\PR[{(y,z)\sim\U[Y\times Z]}]
 {\PR[{x\sim D}]{(x,y,z)\in\Hm}\ge\frac23}\le\frac{\delta}{m}.}}

\prf[\clmref{clm_bjk}]{Assume towards contradiction that that exists some distribution $D_0$ which falsifies the claim.

Corresponding to the statement of the claim is the process of choosing $y\sim\U[M_m]$ and $z=(a,b),$ where $a\sim\U[{[m/2]}]$ and $b\sim\U[\01]$.
The choice is followed by asking what $\PR{(x,y,z)\in\Hm}$ is w.r.t.\ $x\sim D$.
This is equivalent to uniformly choosing two endpoints $i\ne j\in[m]$ and $b\in\01$, followed by asking what the probability is that $x_i\xor x_j=b$ w.r.t.\ $x\sim D$.
Our assumption can be rephrased as
\m[m_dyva]{\PR[{i\ne j\sim\U[{[m]}];b\sim\U[\01]}]
 {\PR[{x\sim D_0}]{x_i\xor x_j=b}\ge\frac23}>\frac{\delta}{m}.}
Define:
\m{C=\set[{i\ne j\in[m];\exists b\in\01:\PR[{x\sim D_0}]{x_i\xor x_j=b}\ge2/3}]
  {\set{i,j}}.}
Since it cannot hold for any $i\ne j$ that both $x_i\xor x_j=0$ and $x_i\xor x_j=1$ occur with probability at least $2/3$, it follows from \bref{m_dyva} that
\f{\sz C\ge\frac{\delta}m(m^2-m)=\delta(m-1).}

Now consider the graph consisting of the edges from $C$.
This graph must contain at least $\sqrt{2\sz C}$ non-isolated vertices, since $v$ vertices give only $(v^2-v)/2<v^2/2$ distinct edges.
Let $C'\sbseq C$ be a forest consisting of a spanning tree for each connected component of this graph.
It must hold that $\sz{C'}\ge\sqrt{\sz C/2}\ge\sqrt{(m-1)\delta/2}.$

Note that the set of uniformly distributed binary random variables \set[\set{i,j}\in C']{x_i\xor x_j} is perfectly independent when $x\sim\U[\01^m]$.
Therefore by entropy subadditivity the entropy loss in $D_0$ is at least 
\m{\sum_{\set{i,j}\in C'}(1-\h[x\sim D_0]{x_i\xor x_j})\ge
 \sz{C'}\dt(1-\h{\beta(2/3)})>
 \frac{\sqrt{(m-1)\delta}}{18},}
where $\beta(2/3)$ denotes the Bernoulli distribution with success probability $2/3$.
Therefore,
\m{\h[D_0]{x}<m-\frac{\sqrt{(m-1)\delta}}{18}=m-\sigma(m),}
which is a contradiction.

The claim follows.}

\ssect{Analyzing $\Hmk$}
Using $k$ parallel copies of the protocol described in \sref{sec_Hm}, 
we obtain a protocol for exact solution of $\Hmk$.
The complexity of the new protocol is $\asO{k\log(m)}$ and it uses $k\log(m)$
EPR pairs.

Now we apply \theoref{theo_dir} together with \clmref{clm_bjk} (note that $\sz{Z_m}=m$ in the case of \Hm).
It follows that
\clm[clm_Hmk_ne]{Any 1-way protocol of communication cost \aso{\frac{k\sqrt m}{\log m}} correctly solves $\Hmk$ with probability $1/2^{\asOm{\frac{k}{\log m}}}$.}

\sssect[ss_lim_ent]{Solving $\Hmk$ with limited entanglement}
In this section we will abuse notation by not distinguishing between a \e{quantum state} and the corresponding density matrix.

The idea of our next argument is the following.
Let $C$ be a 1-way protocol of cost \aso{\frac{k\sqrt m}{\log m}} which start with some entangled state $\rho$ of $e(m,k)$ qubits shared between Alice and Bob and solves $\Hmk$ with probability at least $\delta$.
Denote by $\tau$ the maximally mixed quantum state over $e(m,k)$ qubits.
Consider a modification of the protocol $C$ where instead of $\rho$ we use $\tau$, let us call this new protocol $C'$.

On the one hand, by the laws of quantum mechanics the success probability of $C'$ must be at least $\delta/2^{e(m,k)}$.
On the other hand, because the maximally mixed state over $e(m,k)$ qubits is not entangled, \clmref{clm_Hmk_ne} applies to $C'$.
From that we will derive an upper bound on $\delta$.

Let $\tilde C$ be the measurement-free version of $C$ (i.e., the communication channels are quantum and the output is a quantum state); similarly, let $\tilde C'$ be the measurement-free version of $C'$.
Denote by $U_{x,y}$ the unitary operator corresponding to the action of $\tilde C$ on the shared entangled state when the inputs to Alice and Bob are $x$ and $y$, correspondingly.
In other words if the input pair is $(x,y)$ then $U_{x,y}\rho U^\dagger_{x,y}$ and $U_{x,y}\tau U^\dagger_{x,y}$ are the quantum states obtained after running of $\tilde C$ and $\tilde C'$, correspondingly.

Because $\tau$ is the maximally mixed state of dimension $2^{e(m,k)}$ therefore
$\tau'=\frac{2^{e(m,k)}}{2^{e(m,k)}-1}\llp\tau-\frac1{2^{e(m,k)}}\rrp$ is a
quantum state too.
We can express:
\m{U_{x,y}\tau U^\dagger_{x,y}=
 \frac{2^{e(m,k)}-1}{2^{e(m,k)}}\dt U_{x,y}\tau'U^\dagger_{x,y}+
  \frac1{2^{e(m,k)}}\dt U_{x,y}\rho U^\dagger_{x,y}.}

Let us denote by $\Pi_{x,y}$ the projection of the final state of $\tilde C$
to the subspace of correct answers to $\Hmk(x,y)$.
Our assumption about $C$ can be expressed as
\m{\forall x,y~\tr(\Pi_{x,y}U_{x,y}\rho U^\dagger_{x,y})\ge\delta.}
The success probability of $\tilde C'$ is
\m{\tr(\Pi_{x,y}U_{x,y}\tau U^\dagger_{x,y})\ge
 \frac1{2^{e(m,k)}}\dt\tr(\Pi_{x,y}U_{x,y}\rho U^\dagger_{x,y})\ge
 \frac\delta{2^{e(m,k)}}.}
As mentioned above, \clmref{clm_Hmk_ne} applies to $C'$ and therefore for $m$ sufficiently large,
\f{\frac\delta{2^{e(m,k)}}\in2^{-\asOm{\frac{k}{\log m}}}.}
Therefore, the protocol $C$ can be successful with constant probability only 
if $e(m,k)\in\asOm k.$
This concludes our complexity analysis for \Hmk.

\clm[clm_Hmk]{In the SMP model, $\Hmk$ can be solved exactly by a protocol of cost \asO{k\log(m)} using $k\log(m)$ shared EPR pairs.

In the 1-way communication model with any shared entangled state of \aso{\frac{k}{\log m}} qubits, the communication cost of solving $\Hmk$ with constant-bounded error is \asOm{\frac{k\sqrt m}{\log m}}.}
\theoref{theo_main} follows from this claim.

\sect[sec_2_prov]{Connection to 2-prover proof systems}
Consider the following argument.
Starting with the known 2-prover classical proof system which accepts \NEXP, let us improve its soundness by sequentially repeating the protocol polynomially-many times.
We know that the same proof system is not valid if the provers share entanglement because they can cheat (\cite{CHTW04_Cons}).
However, a straightforward modification of our entanglement bounding approach shows that \e{in order to cheat, the provers require the number of entangled qubits asymptotically close to the number of repetitions}.

Therefore, if the number of shared entangled qubits is bounded by some a priori fixed polynomial in the input length, we can introduce enough repetitions to make any cheating impossible (cf.\ \ssref{ss_lim_ent}).
In combination with the result of \cite{KM03_Qua}, this leads to the following conclusion.
\clm{Let \MIPe\ be the model of 2-prover proof systems in which the provers are allowed to share any entangled state over $e(n)$ qubits, where $n$ is the input length.
If $e(n)\in\poly(n)$ then \MIPe\ can accept a language $L$ if and only if $L\in\NEXP$.}
In other words, the power of \MIPe\ is the same as that of the classical 2-prover proof systems, which is equivalent to \NEXP.

\section{Discussion}
In this paper we solve one of the open question regarding the power of quantum
entanglement in communication complexity:\ we show that no general sublinear
upper bound on the required amount of shared entanglement can be put in the
models of classical communication with either 1-way or simultaneous message
passing.
Can similar results be obtained for other communication models?

In \sref{sec_2_prov} we showed a simple modification of our entanglement
bounding ideas which leads to some nontrivial statement regarding the power of
2-prover proof systems with shared entanglement.
Can we find other applications of our technique outside the domain of
communication complexity? 
Possible applications to other communication complexity models might be
interesting too.
Given that the power of entanglement
is one of the most important longstanding open problems in the area, it is
very tempting to look for other applications of our technique.

There are some more technical questions. 
Can we find more uses for our strong direct product theorem
for relations?
It would be interesting to find applications other than \HMP, though even with
that relation it might probably lead to more results in communication
complexity.
The importance of \HMP\ stems from the fact that this is the only problem we
know today which demonstrates superpolynomial (in fact, exponential)
separation between quantum and classical 1-way communication models.

It would  be also interesting to see whether results similar to ours can be
demonstrated through functional problems, either total or partial (the latter
means that some combinations of $(x,y)$ can never appear in the input).

The last question we would like to mention is whether independent EPR pairs
provide a universal source of entanglement in the contexts of 2-party
communication complexity and 2-prover proof systems.

\subsection*{Acknowledgments}
I thank John Watrous and Ronald de Wolf for helpful discussions.

\bib

\end{document}